\documentclass[]{iopart}

\input{psfig.sty}
\usepackage{iopams}
\begin{document}

\title{Gravitational waves from rotating neutron stars}

\author{D. I. Jones  
\footnote[3]{dij@maths.soton.ac.uk}}

\address{Faculty of Mathematical Studies, University of Southampton, 
Highfield, Southampton, SO17 1BJ, United Kingdom \\}

\begin{abstract}

In this review we examine the dynamics and gravitational wave detectability
of rotating strained neutron stars.  The discussion is divided into two
halves: triaxial stars, and precessing stars.  We summarise recent work on
how crustal strains and magnetic fields can sustain triaxiality, and
suggest that Magnus forces connected with pinned superfluid vortices might
contribute to deformation also.  The conclusions that could be drawn
following the successful gravitational wave detection of a triaxial star
are discussed, and areas requiring further study identified.  The latest
ideas regarding free precession are then outlined, and the recent
suggestion of Middleditch \etal (2000a,b) that the remnant of SN1987A
contains a freely precessing star, spinning-down by gravitational wave
energy loss, is examined critically.  We describe what we would learn about
neutron stars should the gravitational wave detectors prove this hypothesis
to be correct.

\end{abstract}

\maketitle

\section{Introduction}

Broadly speaking, gravitational wave emission from rotating neutron stars
can be divided into two classes: emission due to the normal modes of
oscillation of the fluid core, and emission due to some non-fluid agent
deforming the star, such as the crust or an internal magnetic field.  As a
perusal of journal archives readily reveals, most research into
gravitational wave emission from rotating stars falls into the first class.
The main reason for this is that fluid oscillations can undergo the
Chandrasekhar-Friedman-Schutz (CFS) instability (Friedman \& Schutz
1978a,b), where gravitational radiation reaction amplifies a mode that
propagates forward with respect to the inertial frame, but backward with
respect to the rotating star.  Whether or not such a mode grows to large
amplitude is then determined by the strength of dissipative processes that
tend to sap its energy, such as viscosity of the stellar fluid (see
Lindblom's article in this volume for a review of one such gravitationally
unstable mode - the r-mode).

In this review will concentrate entirely on the relatively neglected second
class of emission, i.e. that due to the neutron star not being a simple
fluid, but instead being able to support some sort of strain.  Such a star
could then be triaxial (i.e. support some sort of `mountain'), and radiate
gravitationally at twice its rotation rate.  There is no gravitational wave
instability for such a star, but equally there will be no viscous processes
within it, allowing a perfectly efficient conversion of kinetic into
gravitational wave energy.  Traditionally, the two candidates for producing
triaxiality have been strains within the crystal lattice of the crust, and
a deformation due to a strong internal magnetic field.  In section
(\ref{sect:tns}) we will describe some recent developments in modelling
these effects, and also describe some new possibilities for producing
deformations, which make use of the pinned component of the neutron star
fluid predicted by glitch modellers.

In addition to triaxiality, a deformed neutron star can radiate
gravitationally by undergoing free precession (sometimes referred to simply
as `wobble').  Such a motion is in the rather unfortunate position of not
being susceptible to the gravitational wave instability of the fluid modes,
and yet still suffering from viscous dissipation within the star. Perhaps
because of this, very little work has been done to investigate the
detectability of freely precessing stars.  In section (\ref{sect:fp}), we
will summarise some recent work on this problem, and comment on the recent
hypothesis of Middleditch \etal (2000a,b) that a gravitationally spinning
down freely precessing pulsar exists in the remnant of supernova 1987A.

\section{Triaxial neutron stars}
\label{sect:tns}

\subsection{Crustal strain}

The most obvious way of deforming the shape of a neutron star from that of
the equivalent rotating fluid is to suppose there is a strain in its solid
crust.  Indeed, very soon after glitches were first observed, a crust-quake
model was proposed (Baym \& Pines 1971, Pines \& Shaham 1972).  The key
idea was that the crust first solidified when the star was spinning very
rapidly, and was therefore very oblate.  As the spin rate gradually
decreased, so did the oblateness, straining the crust until some critical
breaking strain was reached, whereupon the crust cracked.  The consequent
decrease in moment of inertia of the star resulted in the small fractional
increase in spin rate observed as a glitch.

This model soon went out of favour when it was found that not all glitching
pulsars could be described by this model, as the amount of elastic energy
that needed to be stored far exceeded that likely to be available (Lyne \&
Graham-Smith 1998).  Nevertheless, the machinery developed to describe
glitches can be used to estimate how large a non-axisymmetry the crust can
support.  For convenience, we will consider a non-rotating star, with an
axisymmetric deformation whose size we will characterise by the
dimensionless small number $\epsilon$ (this might be the change in radius
along some direction divided by the average radius).  Suppose the crust
would be entirely unstrained if $\epsilon$ took some particular value
$\epsilon_0$.  The energy of such a star can be written as:
\begin{equation}
E(\epsilon) = E_{\rm spherical} 
                     + A\epsilon^2 + B(\epsilon-\epsilon_0)^2,
\end{equation}
where $E_{\rm spherical}$ is the energy the star would have if it were
spherical, $A\epsilon^2$ is the increase in gravitational potential energy
due to the non-sphericity, and $B(\epsilon-\epsilon_0)^2$ is the strain
energy stored in the crust.  Note that the constant $A$ is of order the
total gravitational binding energy of the star, while the constant $B$ is
of the order of the total electrostatic binding energy of the crust.

The actual shape of the star will minimise this energy, so that:
\begin{equation}
\label{eq:epsilon}
\epsilon  =  \frac{B}{A+B} \epsilon_0.
\end{equation}
This is the estimate of crustal deformation we required.  (Strictly, we are
interested in the case where a small strain of the above size is
superimposed upon the shape of a rapidly rotating star, but including the
effects of rotation in our analysis would not change the key result,
i.e. equation \ref{eq:epsilon}).  Fortunately, estimates of $B$ and $A$ are
readily available (see Jones \& Andersson 2001).  The coefficient $B$ is a
property of the crust, and so depends mainly on the rather well known
low-density equation of state.  For a canonical $M = 1.4 M_\odot$ $R =
10^6$ cm neutron star, a value of around $10^{47}$ ergs is found.  The
value will be higher for lower mass stars, as they have thicker crusts.
The coefficient $A$, being of order of the gravitational binding energy of
the whole star, is found to be rather insensitive to the total mass, taking
a value of around $10^{52}$ ergs.  We therefore find:
\begin{equation}
\epsilon \sim 10^{-5} \, \, \epsilon_0.
\end{equation}
for a canonical neutron star.  In words, the electromagnetic forces are
very much weaker than the gravitational ones, so that the deformation
induced by the crust is five orders of magnitude smaller than the `zero
strain' shape of the crust.

This mismatch between the actual shape of the star and its zero strain
shape results in a strain in the crust, of dimensionless size $\sim
|\epsilon - \epsilon_0| \approx |\epsilon_0|$.  This strain cannot exceed 
the crustal breaking strain, $u_{\rm break}$, and so we obtain for the
upper bound on the triaxiality:
\begin{equation}
\epsilon < \frac{B}{A+B} u_{\rm break}.
\end{equation}
Unfortunately, the breaking strain is rather uncertain.  By extrapolation
of breaking strains of terrestrial materials Ruderman (1992) estimates this
to lie in range $10^{-2} \rightarrow 10^{-4}$.  Combining with the
estimates of $A$ and $B$ we finally obtain:
\begin{equation}
\Rightarrow \epsilon_{\rm max} \sim 10^{-7} \rightarrow 10^{-9}.
\end{equation}
In words, we can expect the non-axisymmetric deformations to correspond to
mountains of height in the range $10^{-1} \rightarrow 10^{-3}$ cm.

So much for the bound placed on triaxiality by neutron star structure.
The question remains: Why would a non-axisymmetry be formed in the first
place?  Very little work has been done on this problem.   One possibility
would be that when very young, a hot, entirely fluid neutron  star
undergoes a CFS-type instability, resulting in large violent motions of the
fluid near the surface.  As the star cools, the crust begins to solidify.
If this solidification occurs while the mode is still active, an `imprint'
of the mode might remain in the zero-stress shape of the crust, leaving a
mountain whose structure reflects that of the mode that created it. 

Another rather subtle way of producing a non-axisymmetry was proposed
recently by Bildsten (1998), motivated by the possibility that the low-mass
X-ray binary stars (LMXBs, see Van der Klis 2000) might have their spins
limited by gravitational wave emission.  Bildsten proposed that, providing
a slight non-axisymmetry in the temperature was present, the
compression-induced burning of accreted nuclei will occur in a
non-axisymmetric way, thereby building up an asymmetry in the mass
quadrupole.  This idea was investigated in detail by Ushomirsky, Cutler \&
Bildsten (2000), who solved the combined equations of elasticity and heat
flux for an accreting star.  They found that it was indeed possible to
build up a sufficiently large mass quadrupole for the gravitational
spin-down torque to balance the spin-up accretion torque, provided that
temperature asymmetries at the $5\%$ level were present, and that the
crustal breaking strain is as large as $10^{-2}$, i.e. at the upper end of
the range estimated by Ruderman.

This is a very exciting result for gravitational wave physicists, as one
LMXB in particular, Sco X-1, might be visible to first generation
interferometer detectors, providing sufficiently accurate search templates
are available.  We will discuss gravitational wave detectability of
triaxial stars in section (\ref{sect:gwtri}).

\subsection{Magnetic strain}

The second of the two standard mechanisms for producing triaxiality
concerns neutron star magnetic fields.  Indeed, most of our observational
data on neutron stars comes from the pulsars, whose defining characteristic
is a non-axisymmetry in their magnetic field structure.  The standard
formula for estimating this deformation comes from balancing the total
magnetostatic energy stored by a uniform field $B_{\rm mag}$ within the
star to the gravitational binding energy:
\begin{equation}
\label{eq:mag}
\epsilon \sim \frac{B_{\rm mag}^2 R^3}{GM^2/R} 
         \sim 10^{-12} \left(\frac{B_{\rm mag}}{10^{12} \, \rm G}\right)^2
\end{equation}
for a canonical pulsar.  For typical magnetic fields of order $10^{12}$ G
this is several orders of magnitude smaller than the sorts of deformation
that crustal strain can produce.  However, the size and geometry of the
magnetic field within a neutron star is highly uncertain.  As discussed by
Bonazzola \& Gourgoulhon (1996), if the interior fluid is a type I
superconductor, the magnetic field will be expelled from the core fluid,
and confined to the crustal region.  Such a confinement will result in
internal magnetic fields much larger than the external polar magnetic
field, and can produce deformations several orders larger than formula
(\ref{eq:mag}) would suggest.

\subsection{Magnus strain}

The two forms of deformation considered above have both made use of
electrostatic forces.  However, in the course of attempting to explain
glitches, theorists have realised that other sorts of force may play an
important role in neutron stars.  If the core is a superfluid, it rotates by
forming an array is vortices, with the number per unit area being
proportional to the rotation rate.  About $1 \%$ of the core fluid penetrates
the solid crust.  The vortices in this region are predicted to `pin' to
nuclei in the solid lattice, so that the rotation rate of this fluid is
fixed, even when the crust is spun-up or spun-down.  This creates an
angular velocity difference between the fluid and the vortices, which in
turn results in a Magnus force on the vortices.  This force is
transferred, via the pinning sites, to the crust itself.  

The effect of this Magnus force on the crust has been considered by
Ruderman (1976, 1991, see also Ruderman \etal 1998), who has shown that,
provided that the vortex pinning does not break first, the Magnus force can
crack the crust.  Indeed, Ruderman has used this force to construct a
theory of pulsar magnetic dipole evolution, where the dipole, frozen into
the highly conducting crust, migrates with the fracturing crustal plates
toward the rotational equator of a spinning down star.  A simple estimate
of the size of deformation the Magnus force can produce is straightforward.
Following Ruderman (1991), the force per unit volume on the crystal lattice
is:
\begin{equation}
\label{eq:mf}
\bi{F} = 2  \bomega \times (\bOmega \times \bi{r}) 
                 \rho_{\rm n} f_{\rm pin}
\end{equation}
where $\bomega$ is the angular velocity difference between the superfluid
and the lattice, $\bOmega$ the angular velocity of the star, $\rho_{\rm n}$
the superfluid neutron density and $f_{\rm pin}$ a geometric factor likely
to be of order unity.  The total force on the crust is then of order the
crustal volume $\Delta V$ time this.  The deformation is of the order of
this force divided by the gravitational force at the surface $GM^2/R^2$,
giving
\begin{equation}
\epsilon \sim \frac{2 \omega \Omega R^3 \Delta M}{GM^2},
\end{equation}
where the crust mass $\Delta M \sim \rho_{\rm n} \Delta V$.
Parameterising:
\begin{equation}
\fl 
\epsilon \sim 5 \times 10^{-7} 
         \left(\frac{\omega}{1 \, \rm Hz}\right)
         \left(\frac{f}{\rm kHz}\right) 
         \left(\frac{R}{10^6 \, \rm cm}\right)^3
         \left(\frac{\delta M}{0.01 M_\odot}\right) 
         \left(\frac{1.4 M_\odot}{M}\right)^2 
         \left(\frac{f_{\rm pin}}{1}\right).  
\end{equation}
This is larger than the deformations due to conventional crustal strain,
and so is clearly of interest.

For a spinning-down star (such as a pulsar) $\bomega$ is parallel to
$\bOmega$, and equation (\ref{eq:mf}) shows that the Magnus force is
directed away from the rotation axis, making the crust more oblate than its
rotational equilibrium shape.  Conversely, for a spinning-up star (such as
a LMXB before the compact has has reached spin equilibrium) the Magnus
force tends to make the crust less oblate.  (There must be an equal and
opposite force on the fluid, so the actual change in shape of the star is
not obvious).  However, we are not interested in axi-symmetric
deformations.  We would like to know how the Magnus force might induce
triaxiality.  This question has not, as far as we are aware, been addressed
previously, and is clearly of great interest for gravitational wave
emission.  One (speculative) mechanism would apply if the Bildsten type
asymmetries occur in an accreting system: Any asymmetry in the atomic
number or temperature of nuclei would result in different pinning
strengths.  Certain areas of crust might then pin nuclei perfectly, thereby
feeling the full Magnus force.  Other area might pin nuclei relatively
weakly.  The slippage of the vortices through these areas would then lead
to a somewhat weaker Magnus force being transmitted to the lattice.
Whether or not this could lead to a significant triaxiality is clearly a
problem of interest.

\subsection{Gravitational waves}
\label{sect:gwtri}

\begin{figure}[h]
   \centerline{ \psfig{file=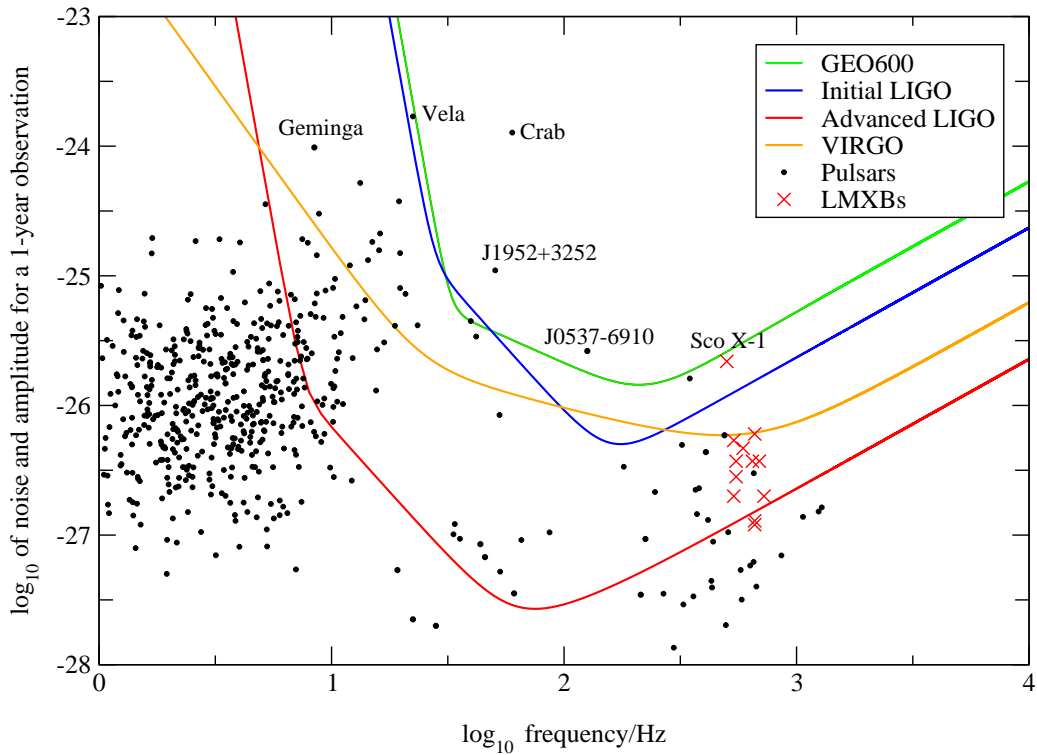,angle=270,width=16cm} }
   \caption{Upper bound on $h$ from spin-down rates for pulsars and
   accretion luminosity for LMXBs.  Pulsar data from Princeton catalogue
   and Lorimer (2001), LMXB data from Bildsten (1998).}
\label{fig:gwtri}
\end{figure}

Having reviewed the physics behind triaxiality, we now reach our intended
destination: a discussion of gravitational wave detection.  Figure
\ref{fig:gwtri} is a modern version of the classic `upper bound on the
gravitational wave amplitude' diagram.  In the figure we plot gravitational
wave amplitudes and detector noises verses frequency.  Each black dot
represents a pulsar of known period and period derivative.  These two
pieces of data, together with an estimate of the distance from Earth and
the star's moment of inertia (assumed to be $10^{45} \rm \, g \, cm^2$ for
all) allows an estimate of the gravitational wave amplitude, \emph{
assuming all of the spin-down energy is converted to gravitational wave
energy}.  A similar diagram could have been drawn at any point over the
last twenty years.

However, the crosses represent a rather recent development---the LMXB
systems.  As described by Papaloziou \& Pringle (1978), and calculated for
the new LMXB data by Bildsten (1998), a measurement of the distance,
electromagnetic luminosity, and spin rate of an accreting star allows a
calculation of a gravitational wave amplitude.  The key assumption here is
that the accreting plasma exerts a torque at the stellar surface
appropriate to it being in a Keplerian orbit just prior to accretion.
(Note that there is some uncertainty in the exact spin rates of these
objects, connected with the interpretation of quasi-periodic oscillations
in their electromagnetic fluxes---see van der Klis 2000).

Clearly, the wave amplitudes in the figure represent upper bounds, probably
rather optimistic ones in the case of the pulsars, whose spin-down energy
budget probably involves electromagnetic energy losses in addition to
gravitational ones.  Nevertheless,  we should be ready, in the event of a
successful gravitational wave detection, to try and draw conclusions
regarding neutron star structure.  

The first thing to note is that there are two separate groups of stars in
the figure that lie above or close to the first-generation noise curves.
At the low frequency end we have the young pulsars (e.g. the Crab and
Vela), while at the high frequency end we have the old millisecond pulsars
and the LMXBs.  A successful detection of a triaxial star would lead to very
different conclusions being drawn, depending upon whether it is a low or
high frequency source that is observed.  

Working backwards, the degree of triaxiality required to explain the
gravitational wave spin-down of a typical young pulsar in the diagram is of
order $10^{-3}$, approximately four orders of magnitude larger than the
largest triaxialities discussed previously.  If such a gravitational wave
signal were to be observed, we would be forced to conclude that our picture
of neutron star interiors is seriously lacking.  If magnetic fields were
responsible, an internal field in excess of $10^{16}$ G would be required
(equation \ref{eq:mag}), around four orders of magnitude larger than the
external polar field.  Magnus forces are certainly too weak to produce such
a triaxiality.  The remaining possibility might be that our knowledge of
the high density equation of state is poor, and that a solid, highly
rigid core exists within the star, capable of supporting a much larger
triaxiality than the relatively low density crust.  Such a result would be
very exciting, and full-fill the promise that gravitational wave astronomy
can probe deeper into neutron star interiors than electromagnetic
observations.

Observation of a gravitational wave signal at or close to the strengths
indicated from one of the high frequency sources would lead to very
different conclusions.  The degree of triaxiality required to explain this
strength of emission is around $10^{-8}$, comfortably within the range of
$\epsilon$ values discussed previously.  In terms of our theoretical
expectations, these high frequency sources seem much better bets than the
low frequency ones, an expectation that may guide gravitational wave data
analysts in their searches.  However, this creates a new problem: If such
a source were to be observed, how could we decide which of the three
deformation mechanisms described above (if any) is producing the
deformation?    

One partial solution suggests itself, based on the fact that if it is a
strong magnetic field that is producing the triaxiality, the quadrupole
moment (and therefore the gravitational wave phase and amplitude) should
not be significantly affected by a glitch, whereas if either crustal strain
or Magnus forces are responsible, the gravitational wave amplitude and
possibly the phase should both change abruptly.  However, in over 30 years
of observation, a millisecond pulsar has not yet been observed to glitch
(Lyne \& Graham-Smith 1998), while the spin rates of the LMXBs are not
known accurately or monitored frequently (van der Klis 2000), so further
distinguishing features need be found.  Clearly, this is a tricky problem,
and one which seems not to have been investigated previously.
Nevertheless, its resolution is required if we are to ever to fulfil our
promise of using gravitational wave data to learn about the interiors of
these stars.

\section{Free precession}
\label{sect:fp}

\begin{figure}
\begin{center}
\begin{picture}(60,90)(-20,0)

\thicklines
\put(0,0){\vector(0,1){80}}

\put(0,0){\vector(-2,3){40}}

\put(0,83){$\bf{J}$}

\put(0,0){\vector(1,4){13}}

\thinlines

\qbezier(-5.547,8.321)(-2.8,10.6)(0,10)

\put(13,55){$\bf{\Omega}$}

\put(-40,63){$\bf{n}$}

\put(-4.8,11){$\bf{\theta}$}

\end{picture}

   \caption{The geometry of free precession.  The symmetry axis of the
   deformation lies along $\bf{n}$, which moves in a cone of half-angle
   $\theta$ around the angular momentum vector $\bf J$.  The angular
   velocity vector, denoted by $\bOmega$, lies in this plane also.}
\label{fig:fp}
\end{center}
\end{figure}
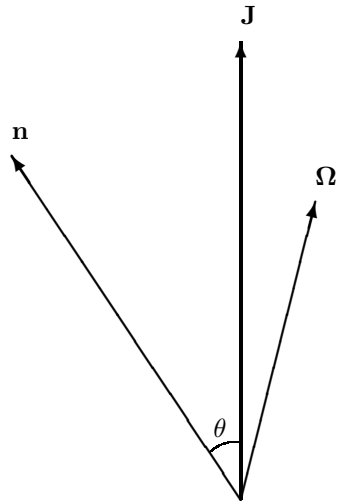

Free precession is the most general motion of a rigid body (Landau \&
Lifshitz 1976).  For an axisymmetric rigid body, the motion consists of the
symmetry axis moving in a cone of half-angle $\theta$ (known sometimes as
the `wobble angle') about the fixed angular momentum vector $\bf{J}$.  The
angular velocity vector $\bOmega$ lies in this plane also, as illustrated
in figure \ref{fig:fp}.  The whole plane rotates around $\bf J$ at a rate
that we will call $\dot \phi$.  In addition to this rotation, there is a
superimposed slow rotation of the star about $\bf n$, at a rate $\dot \psi$
known as the body-frame precession frequency.  The two rotation rates are
related by:
\begin{equation}
\label{eq:psidot}
\dot \psi = -\frac{\Delta I}{I_0} \dot \phi
\end{equation}
where $\Delta I$ is the non-spherical piece of the moment of inertia tensor
symmetric about $\bf n$, and $I_0$ is the total moment of inertia along
this axis.  Free precession was first discussed as a source of
gravitational radiation by Zimmermann (1978) and Zimmerman \& Szedenits
(1979), who showed that mass quadrupole gravitational radiation was
produced at the frequencies $\dot \phi$ and $2 \dot \phi$.  If the star
radiates electromagnetically as a pulsar, the pulsation rate is given by a
rather complicated time-varying combination of $\dot \phi$ and $\dot \psi$
(Jones \& Andersson 2000).

Of course real neutron stars are not rigid---they consist of a thin elastic
shell containing a superfluid core, possibly with part of the fluid pinned
to the crust.  With the obvious exception of the superfluidity and pinning,
the Earth itself has a similar structure, so the machinery that
geophysicists have evolved to describe the Earth's free precession (known as
the `Chandler wobble') can be used to begin to describe neutron stars
(Pines \& Shaham 1972).  One important point is that, because the crust is
elastic, during the course of one free precession period its shape changes,
corresponding to the movement of the `centrifugal bulge', symmetric about
$\bOmega$, in a cone around $\bf n$.  This places a limit on the wobble
angle given by:
\begin{equation}
\label{eq:thetamax}
\theta_{\rm max} \approx \frac{u_{\rm break}}{\epsilon_{\rm rotation}} 
       \approx 0.45 \left(\frac{100 \, \rm Hz}{f}\right)^2            
       \frac{u_{\rm break}}{10^{-3}}.
\end{equation}
where $\epsilon_{\rm rotation} \approx \Omega^2R ^3 /GM$, is the
(dimensionless) size of the centrifugal deformation.  The effects of
pinning can be added relatively straightforwardly (Shaham 1977).  The final
picture that emerges is that the basic geometry described above applies
even to realistic neutron stars, although the free precession timescale
depends upon the internal structure in a complicated way (Jones \&
Andersson 2000).

Gravitational wave emission from realistic precessing stars was examined by
Alpar \& Pines (1985), who pointed out that for a long-lived freely
precessing star to exist, some mechanism must be found to continually
excite the wobble motion, as dissipative processes within the star tend to
damp the motion.  The magnitude of the internal dissipation  was
estimated by Alpar \& Sauls (1988), who found that, in the absence of any
excitation mechanism, free precession is damped in between $400$ and $10^4$
free precession periods.

There have been three very recent developments in this field.  First of
all, Stairs, Lyne \& Shemar (2000) have reported evidence of a freely
precessing pulsar, identified by careful analysis of the electromagnetic
pulse timing data.  This pulsar spins very slowly, and so is not of
gravitational wave interest, except inasmuch as it allows us to test our
understanding of the physics of precession.  As noted by Jones \& Andersson
(2000), the timing data were consistent with theoretical expectations,
\emph{provided that none (or virtually none) of the superfluid is pinned to the
crust}, seemingly in conflict with the standard glitch model of pulsar
behaviour.  A resolution to this seeming paradox was supplied by Link \&
Cutler(2001), who showed that the Magnus forces induced by the precessional
motion might be sufficient to break the vortex pinning, provided that the
pinning is sufficiently weak.  In this way, the standard pulsar model could
continue to explain glitches in non-precessing pulsars, while still
accounting for the observations of Stairs \etal.

Secondly, and of most interest to us, Jones \& Andersson (2001) have
recently undertaken a systematic study of the likely free precession
gravitational wave amplitudes likely to exist in Nature (see also Jones
2001).  The problem is one of finding sufficiently strong wobble-pumping
mechanisms to counteract the dissipation of precessional energy within the
star.  In this study, many different pumping mechanisms were considered,
including accretion torques, electromagnetic torques on pulsars, wobble
induced by the violent birth of a neutron star in a supernova, and near
collisions between a neutron star and another star.  In all cases it was
found that the dissipation of precessional energy within the neutron star
limited the gravitational wave amplitude to a level undetectable even by a
second generation laser interferometer.  These conclusions were reached
assumed that no superfluid was pinned to the crust, is agreement with the
Stairs et al (2000) observation and the Link \& Cutler (2001) pinning
model.  If a pinned component is included, the internal dissipation rate is
even higher, reinforcing this pessimistic conclusion: freely precessing
neutron stars don't seem to be good sources of gravitational radiation.

However, the third recent development in this field is a possible
observational of a \emph{gravitationally spinning-down pulsar}, reported by
Middleditch \etal (2000a,b).  The pulsar in question was found in the
remnant of supernova 1987A, and was observed only intermittently between
1992 and 1997, with no reported sightings since.  The free precession
period $2 \pi / \dot \psi$ seemed to vary, spanning a range of 935s to
1430s.

The main piece of evidence Middleditch et al.\ cite to support the
gravitational wave spin-down hypothesis is that the frequency derivative
$\dot f$ correlates rather well with the inverse square of the long
modulation period $P_{\rm fp}$, with $\dot \Omega$ varying by a factor of
$\sim 4$, while $P_{\rm fp}$ varies by a factor of 2.  The spin-down rate
of a precessing pulsar is given by (Cutler \& Jones 2000):
\begin{equation}
\label{eq:Omegadot}
\dot \Omega = \frac{32G}{5c^5} \Omega^5 
              \frac{(\Delta I)^2}{I_{\rm star}}
              \sin^2 \theta (\cos^2 \theta + 16 \sin^2 \theta)
\end{equation}
Combining equations (\ref{eq:psidot}) and (\ref{eq:Omegadot}) we see that
just such a correlation is to be expected, \emph{provided that the wobble
angle remains roughly constant}.  However, this is a somewhat curious state
of affairs: it is difficult to imagine how the deformation in the moment of
inertia tensor $\Delta I$ varies by a factor of two, while maintaining a
roughly constant orientation with respect to the angular momentum vector.

The hypothesis of gravitational wave spin-down via free precession can be
tested further.  Note that $\Omega$, $\dot \Omega$, the ratio $\Delta I /
I_0$ can all be extracted from the observational data, while $I_0$
estimated as $10^{43} \, \rm g \, cm^2$ if just the crust participates in
the free precession, and $I_{\rm star} \sim 10^{45} \, \rm g \, cm^2 $.
Then equation (\ref{eq:Omegadot}) can be inverted to give $\theta =
25^\circ$.  Using equation (\ref{eq:thetamax}) we see that this requires a
crustal breaking strain of approximately $2 \times 10^{-2}$, higher then
even the most optimistic estimate of Ruderman (1992).  This difficulty is
relieved somewhat if the whole star is assumed to participate in the free
precession; then equation (\ref{eq:Omegadot}) gives $\theta = 9^\circ$, and
equation (\ref{eq:thetamax}) gives a crustal breaking strain of $8 \times
10^{-3}$, a somewhat more plausible value.  In summary, if this pulsar
really is spinning-down due to gravitational wave energy loss, the breaking
strain of the crust must be very high, particularly if only the crust is
participating in the free precession.  By coincidence(?), the breaking
strain required, approximately $10^{-2}$, is close to the value required by
Ushomisrsky, Cutler \& Bildsten (2000)to support the triaxiality of the LMXBs.

To sum up, if gravitational wave observations were to confirm the free
precession hypothesis for this pulsar, we will have learnt three important
facts about neutron stars.  Firstly, their deformations can vary
significantly over rather short timescales (of order a few years); their
crusts are rather strong, with breaking strains at the upper end of
theoretical estimates; and finally that they can be set into free
precession during, or soon after, birth.  Clearly, when the laser
interferometers reach the necessary sensitivity, a search for this
precession candidate will be of great interest.

\section{Conclusions}

In this review we have examined our current understanding of the dynamics
and gravitational wave detectability of rotating strained neutron stars.
We divided our discussion into two halves: triaxial stars, and precessing
stars.  Both fields, having been rather quiet for some years, have enjoyed
an injection of interest recently: The observations of spin frequencies in
the LMXBs exciting interest into the triaxial problem, and observations
from Stairs \etal (2000) and Middleditch \etal (2000a,b) exciting interest
in the free precession case.

For triaxial stars we noted that, for first generation gravitational wave
detectors at least, an observation of gravitational wave spin-down from one
of the young pulsars would require us to profoundly rethink our picture of
neutron star interiors.  A successful detection of gravitational waves from
a millisecond pulsar or LMXB would be less spectacular, as this level of
emission is consistent with current theories of stellar deformation.
However, the problem of deciding just what was the mechanism responsible
for the deformation has not yet been addressed, and stands in the way of
our fulfilling our promise of using gravitational wave observations to
probe neutron star interiors.  To complicate the issue further, we have
speculated that, in addition to the traditional crustal strain and magnetic
field-induced deformation, triaxiality might be produced by the Magnus
force on superfluid vortices pinned to the crust.  A study of how such a
force might set up a non-axisymmetry in a pulsar is clearly of great
interest.

For freely precessing stars, the recent study of Jones \& Andersson (2001)
suggests that precessing stars do not seem to be a promising source of
gravitational radiation.  The reason for this is that dissipative forces
within the stellar interior rapidly damp the precessional motion.  However,
Middleditch \etal (2000a,b) have recently presented evidence of a pulsar
within the remnant of supernova 1987A, and have suggested that it is
spinning-down due to gravitational waves generated by free precession.  We
have investigated this hypothesis, and found that if correct we would learn
a number of surprising things about neutron stars, not least that their
deformations can vary in size significantly on a timescale of a few years.

To sum up, after having been rather quiet for sometime, the field of
gravitational wave emission from strained neutron stars has heated up.
Although still a poor cousin of the unstable fluid modes and binary
inspiral, the types of source considered here will be high on the target
lists of gravitational wave data analysts.  As we hope this review makes
clear, much could be learnt from a successful detection, but more work still
need to be done to turn observations into constrains on the physics, and
thereby turn gravitational wave astronomy into gravitational wave
astrophysics.

\ack

I would like to thank the organisers for arranging an excellent
conference.  This work was supported by PPARC grant PPA/G/1998/00606.

\section*{References} 
\begin{harvard}

\item
Alpar A., Pines D., 1985, {\em Nature} {\bf 314} 334

\item
Alpar A., Sauls J. A., 1988, {\em Ap. J.} {\bf 327} 723

\item
Baym B., Pines D., 1971, {\em Annals of Physics} {\bf 66 } 816

\item
Bildsten L., 1998, {\em Ap. J.} {\bf 501} L89

\item
Bonazzola s., Gourgoulhon E., 1996, {\em Astron. Astrophys.} {\bf 312} 675

\item
Cutler C., Jones D. I., 2000, {\em Phys. Rev. D} {\bf 63} 024002

\item
Friedman J. L., Schutz B. F., 1978a, {\em Ap. J.} {\bf 221} 937

\item
Friedman J. L., Schutz B. F., 1978b, {\em Ap. J.} {\bf 222} 281

\item
Jones D., 2001, in Ferrari V., Miller J., Rezzolla L., eds, Gravitational
Waves: A Challenge to Theoretical Astrophysics, Abdus Salam International
Centre for Theoretical Physics

\item
Jones D. I., Andersson N., 2001, {\em MNRAS} {\bf 324} 811

\item
Landau L. D., Lifshitz E. M., 1976, Mechanics, 3rd Edition.
Butterworth-Heinemann Ltd.

\item
Link B., Cutler C., 2001, Submitted to MNRAS

\item
Lyne A. G., Graham-Smith F., 1998, Pulsar Astronomy.  Cambridge University
Press

\item
Middleditch J., et al.\, 2000a, {\em New Astronomy} {\bf 5} no.\ 5, 243

\item
Middleditch J., et al.\, 2000b, astro-ph/0010044

\item
Lorimer, D., 2001, Liv. Rev. Rel. 4

\item
Papaloizou J., Pringle J. E., 1978, {\em MNRAS} {\bf 184} 501

\item
Pines D., Shaham J., 1972, {\em Phys. Earth and Planet. Interiors} 
{\bf 6} 103

\item
Ruderman M., 1976, {\em Ap J.} {\bf 203} 213

\item
Ruderman M., 1991, {\em Ap J.} {\bf 366} 261

\item
Ruderman M., 1992,
In {Structure and Evolution of Neutron Stars} p.353

\item
Ruderman M.,  Tianhua Z., Kaiyou C., 1998, {Ap. J.} {\em 492} 267

\item
Shaham J., 1977, {\em Ap. J.} {\bf 214} 251

\item
Stairs I. H., Lyne A. G., Shemar S. L., 2000, {\em Nature} {\bf 406} 484

\item
Ushomirsky G., Cutler C., Bildsten L., 2000, {\em MNRAS} {\bf 319} 902

\item
van der Klis M., 2000, {\em Annual Rev. Astron. and Astrophys.} {\bf 38}
717

\item
Zimmermann M., 1978, {\em Nature } {\bf 271 } 524

\item
Zimmermann M., Szedenits Jr. E., 1979, {\em Phys. Rev. D } {\bf 20 } 351

\end{harvard}

\end{document}